\documentclass{aa}
\usepackage{graphicx, natbib, epsfig, amsmath, amssymb, mathrsfs, txfonts}
\DeclareGraphicsExtensions{.eps, .jpg, .ps}
\usepackage{epstopdf}
\bibpunct{(}{)}{;}{a}{}{,}
\usepackage{color}

\begin{document}

\title{Laser-guide-stars used for cophasing broad capture ranges
}
\author{P. Martinez and P. Janin-Potiron}
\institute{Universit\'e C\^ote d\'{}Azur, Observatoire de la C\^ote d\'{}Azur,  CNRS, Laboratoire Lagrange, Parc Valrose, B\^at. Fizeau, 06108 Nice, France}
\offprints{patrice.martinez@oca.eu}

\abstract
{Segmented primary mirrors are indispensable to master the steady increase in spatial resolution. Phasing optics systems must reduce segment misalignments to guarantee the high optical quality required for astronomical science programs.} 
{Modern telescopes routinely use adaptive optics systems to compensate for the atmosphere and use laser-guide-stars{} to create artificial stars as bright references in the field of observation.
Because multiple laser-guide-star adaptive optics are being implemented in all major observatories, we propose to use man-made stars not only for adaptive optics, but for phasing optics.}
{We propose a method called the doublet-wavelength coherence technique (DWCT), exploiting the D lines of sodium in the mesosphere using laser guide-stars. The signal coherence properties are then used.}
{The DWCT capture range exceeds current abilities by a factor of 100. It represents a change in paradigm by improving the phasing optics capture range from micrometric to millimetric. It thereby potentially eliminates the need of a man-made mechanical pre-phasing step.}
{Extremely large telescopes require hundreds of segments, several
of which need to be substituted on a daily basis to be
recoated. The DWCT relaxes mechanical integration requirements and speeds up integration and re-integration process.} 
\authorrunning{Martinez}
\maketitle

\section{Introduction}
Adaptive optics (AO) enables obtaining diffraction-limited imaging provided that a guiding light-source is used as a reference for the wavefront correction. Bright natural stars are required for
this, but the sky coverage is unfavorably low, in particular in the visible wavelength range. To overcome this limitation, artificial laser-guide-stars (LGSs) have been proposed in the 1980s \citep{LGS85} to create a source that lies above the turbulent layers of the atmosphere. 
LGS-based AO systems are now routinely used in all leading observatories (e.g., Keck, Gemini, ESO/VLT, and Lick), and multiple LGS AO systems are intensively employed, as demonstrated by the four LGSs that simultaneously shone above the Unit 4  telescope (UT4) at ESO's VLT in early May 2016. 

Different types of lasers are suitable for AO applications, but the most recent LGS systems make use of mesospheric scattering of a sodium laser. 
The principle of the sodium guide star is to tune the wavelength of the laser radiation to a resonance of sodium atoms at 589\,nm (D2 line). Sodium atoms, naturally present in the mesosphere at an altitude of 95\,km (on average), will absorb laser light and subsequently emit fluorescence at the same wavelength. The intensity of the created star depends on the amount of sodium atoms present at the altitude of the mesospheric sodium layer. In this context, various astronomical and atmospherical studies have been conducted over the past decades to understand the temporal and spatial characteristics of the atmosphere sodium layer \citep[e.g., ][]{GE98, P1999, A2000, SLANGER2005, PLANE2012}. 

The future of ground-based astronomy in the next decades is bound to the next generation of extremely large telescopes (ELTs).   
ELTs represent a major change in dimension, wavefront control strategies, and execution time.
The wavefront control of ELTs includes three main systems: the aforementioned AO system that corrects for the atmospheric turbulence, active optics that correct for misalignments and deformations
of the telescope adaptive mirror, and phasing optics that correct for the misalignment of individual segments of the primary mirror. Generally, cophasing is a three-step procedure to correct for the initial misalignment imposed by mechanical structure constraints to the final alignment: (1) a mechanical pre-phasing step using hand-held optical tools, (2) a coarse-phasing step using a dedicated phasing sensor, and (3) a fine-phasing step at high precision using a dedicated phasing sensor. 
Step (1) will be critical for ELTs because their primary mirrors require hundreds of segments, several of which will frequently need to be substituted for recoating (the reflectivity coating lifetime is generally limited to 18 months). 
In this context, mechanical integration and step (1) must reduce the piston error to the capture range of actual optical cophasing sensors. 
Enlarging this capture range is indispensable to relax mechanical integration requirements and to speed up integration and re-integration process before the observing run. 

Conventional coarse-phasing techniques use multiple wavelength or coherence methods \citep[e.g., ][]{CHANAN98, VIGAN11} to increase the capture range of the phasing sensor, which is limited by the so-called $\pi$-ambiguity otherwise.
These methods successfully enlarge the initial capture range of a cophasing sensor to $\sim10\, \mu m$ $rms$. With the method
we propose, segments of a telescope can reliably be phased from a $rms$ piston error of more than $1000\, \mu m$. Our method is called the doublet-wavelength coherence technic (DWCT) and
does not require any hardware, except for the LGSs offered by the telescope. 
The DWCT is a new solution based on multiple LGSs lasing at different wavelengths through the analysis of the coherence signature in the resulting cophasing sensor signal. 
Because it exceeds the actual extended capture range of the cophasing sensor by a factor of 100, it might eliminate the need of the man-made mechanical pre-phasing step (1) along with inherent stability and compatibility independent of the phasing sensor type. 

In the following sections, the general background of phasing optics 
is introduced, and conventional coarse measurement techniques (multiwavelength and coherence methods) are presented for pedagogical reasons and because they are considered the basis for this work. Then the theory of the proposed DWCT and its advantages and limitations are discussed.

\label{introduction}

\section{Cophasing background}
\label{section1}
The initial difference in the optical path between segments before cophasing is turned on typically is about 100 \,$\mu m.$ This is imposed by the mechanical structure tolerance, step (1), and the residual after step (2) is expected to be of $\sim100 \, nm$, which is ultimately reduced to $\sim10\, nm$ after step (3).  This gives rough numbers of misalignment levels and/or expected residuals that a phasing sensor has to cope with. On the other hand, the capture range of a cophasing sensor is strictly limited, and it characterizes its ability to detect misalignment phases over a given range of wavelengths. The capture range can be either defined in monochromatic (intrinsic capture range) or broadband light (extended capture range), with remarkable differences.  

In monochromatic light, a phasing sensor signal ($\mathrm{S}$) is proportional to a sinusoidal function of the wavefront step height $\epsilon$ such as 
\begin{equation}
\mathrm{S}(\epsilon) = \mathrm{A} + \mathrm{C} \sin \left(\mathrm{k_0} \epsilon  \right),
\label{Eq1}
\end{equation}
\noindent where $\mathrm{k_0}=  2\pi/ \lambda_0$ is the wave number, $\lambda_0$ is the working wavelength (assumed monochromatic), $\mathrm{A}$ and $\mathrm{C}$ are constants, and $\epsilon/2$ is the physical mirror step height. Equation \ref{Eq1} trivially introduces the well-known problem of any cophasing sensor in the monochromatic regime, the so-called $\lambda$-ambiguity. For simplicity reasons and without loss of generality, we omit the constant $\mathrm{A}$ in the next developments. 

\subsection{ $\lambda$-ambiguity problem}
Because the estimated signal (Eq. \ref{Eq1}) is $\lambda$-periodic, the capture range is strictly limited to the non-ambiguous range of the sine function. This problem appears in all phasing sensors operating in monochromatic light, and limits the capture range to $\pm$ $\lambda/2$, the so-called $\pi$-ambiguity. Within the capture range, the piston can be corrected unambiguously. But if the edge piston error between two segments is larger than $\lambda/2$, phasing the two segments with zero phase error cannot be achieved, only to the nearest integer ($n$) multiple of the wavelength. 
As a consequence, instead of being phased to zero, various segments will be phased with an ambiguity left of $n \lambda$.
To overcome this problem and to increase the capture range of phasing sensors, conventional techniques based on multiwavelength or coherence analysis (broadband light) have been proposed. 

\subsection{Solving the $\lambda$-ambiguity}
\subsubsection{Multiwavelength method}
The multiwavelength method performs successive phasing at two (or more) different and discrete wavelengths using two optical filters ($\lambda_1$ and $\lambda_2$) to determine the ambiguity $n$ for every segment. 
The method is straightforward because the piston difference between the two phased positions successively obtained at $\lambda_1$ and $\lambda_2$ is related to the ambiguity $n$ and the two wavelengths. 
The capture range in the multiwavelength scheme is limited to $\pm \Lambda/2$, where $\Lambda$ is the synthetic wavelength equal to 
\begin{equation}
\Lambda = \frac{\left( \lambda_1 + \lambda_2 \right)^{2}}{2 \left( \lambda_1 - \lambda_2 \right)}.
\label{Eq2}
\end{equation}
This method is still limited by the periodic nature of the signal functions at the two wavelengths, however. This imposes a rigid condition to the selection of the two wavelengths: the ambiguity $n$ must be identical for the two selected wavelengths, thus limiting the achievable capture range to some
extent. This method has been successfully demonstrated with several cophasing sensors \citep[][]{PINNA2006, VIGAN11, SURDEJTHESIS}. 

\subsubsection{Coherence method}
\label{coherence}
The coherence technique uses the coherence length in broadband light of an optical filter with a finite spectral bandwidth $\delta \lambda_F$. 
This method takes advantage of the wavelength-dependent signature in the Fourier space to extend the capture range beyond the $\pi$-ambiguity restriction.
This signature exhibits a coherence envelope, whose characteristic scale is the so-called coherence length of the optical filter and is given by 
\begin{equation}
\mathrm{L_c} = \frac{\lambda_0^2}{\delta \lambda_\mathrm{F}},
\label{Eq3}
\end{equation}
\noindent where $\lambda$ is the central wavelength. 
In this context, Eq. \ref{Eq1} still applies in broadband light except that $\mathrm{C}$ is no longer a constant but a function of $\mathrm{k}$.
By assuming a Gaussian bandpass in $\mathrm{k}$ centered on $\mathrm{k_0}$, we can write the filter profile function as  
\begin{equation}
\mathrm{P}(\mathrm{k}) =  \frac{1}{\sqrt{2 \pi \sigma^2}} \times \exp{\frac{-\left( \mathrm{k} - \mathrm{k}_0 \right)^2}{2 \sigma^2}},
\label{Eq4}
\end{equation}
\noindent where $\sigma$ is given by
\begin{equation}
\sigma =   \frac{2 \pi}{\sqrt{8 \ln 2}}\times  \frac{\delta \lambda_F}{\mathrm{\lambda_0^2}}, 
\label{Eq6}
\end{equation}
\noindent and using Eq. \ref{Eq3}, $\sigma$ can be expressed as function of the coherence length such as
\begin{equation}
\sigma =    \frac{2 \pi}{\sqrt{8 \ln 2}}\times  \frac{1}{\mathrm{L_c}},
\label{Eq6b}
\end{equation}
\noindent and thus summing over all the wavelengths, that is, integrating over $\mathrm{k}$ replacing $C$ by $P(k)$, the broadband sensor signal can finally be approximated to
\begin{equation}
\mathrm{S}(\epsilon) \approx \sigma \sqrt{\frac{2}{\pi}} \times \exp{ \left( \frac{-\left( \sigma \epsilon \right)^2}{2} \right)} \times  \sin \left( \frac{2 \pi}{\lambda_0} \epsilon  \right).
\label{Eq7}
\end{equation}
In broadband light, the sensor signal is still proportional to a sine function of the wavefront step height, but modulated by a Gaussian distribution. Thus the periodic nature of the sine is no longer a problem because it decays as the wavefront step height error increases. This corresponds to the coherence signature that can be exploited. 

When individual segments are successively poked by varying heights
in the wavefront step, the estimated signal will describe a sinusoidal function ($\lambda$-periodic) modulated by the Fourier transform of the optical filter profile function (the coherence envelope). Analyzing the variation in the measured coherence function enables
us to identify and correct for the $n$ ambiguity for every segment. 
The capture range of the phasing sensor is then enhanced to $\pm$ $\mathrm{L_c}$. 
Because the characteristic scale of the measured coherence function is the coherence length of the optical filter, the selection of the optical filter bandpass ($\delta \lambda_F$) is critical to match the range of piston errors that is expected to occur
at the expense of precision, and vice versa.
This method has been validated with several cophasing sensors with carefully selected bandpasses \citep[e.g., ][]{CHANAN98, SURDEJTHESIS}. 

\subsection{Discussion}
Using these techniques, the capture range can be extended from the half-wave limitation to a few tens of $\mu$m (typically up to 10 $\mu m$) upon the method selected \citep[e.g., ][]{CHANAN98, VIGAN11, SURDEJTHESIS}. Another method exploiting the coherence signature introduced by a liquid crystal tunable filter \citep{Bonaglia08} shows an even greater capture range (50 $\mu m$). However relevant and efficient, these techniques have the disadvantage that they
require special hardware and use a time-consuming iterative process by repeating the calibration process for multiple filters or wavelengths. 
The coherence method must start with a narrow bandwidth filter that covers a broad range of pistons at the expense of reduced accuracy, and the filter bandwidth is increased iteratively to improve the precision at the best-fit alignment. 
In addition, the achieved capture range does not mean that the optical phasing operation can be omitted from the mechanical pre-phased step (1). It is still worth exploring new methods that will ultimately not be affected by these constraints. 

\section{Doublet-wavelength coherence technique}
\label{section2}
The DWCT exploits a coherence signature using two wavelengths
simultaneously. For the sake of generality, it is assumed in the following that the two lines are non-equal in intensity.

\subsection{Analytical treatment}
We consider two simultaneous and discrete lines at wavelength $\lambda_1$ and $\lambda_2$, with associated intensity $\mathrm{I_1}$ and $\mathrm{I_2}$. For simplicity reasons, we first assume that the spectral distributions of the two lines are described by Dirac functions.  
We note by $\mu$ the ratio of the two line intensities ($\mu=\mathrm{I_2}/\mathrm{I_1} \ne 1$). 
Again, Eq. \ref{Eq1} still applies in this case, but $\mathrm{C}$ is a function of $\mathrm{k}$, and the sensor signal is still proportional to a sine function of the wavefront step height, but modulated by a cosine distribution. The signal sensor is given by
\begin{equation}
\mathrm{S}(\epsilon) \approx \mathrm{I_1} \left( 1 + \mu \right)  \times \left[ 1 +\cos \left( \frac{ \pi \Delta \lambda}{\lambda_c^2} \epsilon  \right) \times  \sin \left( \frac{2 \pi}{\lambda_c} \epsilon  \right) \right],
\label{Eq8}
\end{equation}
\noindent where $\Delta \lambda$ is the wavelength separation between the two lines ($\Delta \lambda = \lambda_2 - \lambda_1$)
, and $\lambda_c = \left(\lambda_1 + \lambda_2 \right)/2$ is the central wavelength. 
The modulation introduced by the cosine function is the expression of the coherence signature in the particular case of a doublet-line system. 
Nevertheless, since the cosine function is $\Delta \lambda$-periodic, it does not solve the $\pi$-ambiguity problem. However, the spectral distribution of the two lines is not described by Dirac but by
Gaussian functions  following Eq. \ref{Eq4}, but where $\sigma=\sigma_c$ such that
\begin{equation}
\sigma_c =   \frac{2 \pi}{\sqrt{8 \ln 2}}\times  \frac{ \delta \lambda}{\mathrm{\lambda_c^2}}, 
\label{Eq9}
\end{equation}
\noindent where $\delta \lambda$ is the line width. 
Assuming Eq. \ref{Eq4} for the line profiles, it modifies Eq. \ref{Eq8} to
\begin{equation}
\begin{split}
\mathrm{S}(\epsilon) \approx \mathrm{I_1} \left( 1 + \mu \right) \times \\ \left[ 1 + \sigma_c \sqrt{\frac{2}{\pi}} \times \exp{ \left( \frac{- (\sigma_c \epsilon)^2}{2} \right)} \times \cos \left( \frac{ \pi \Delta \lambda}{\lambda_c^2} \epsilon  \right) \times  
\sin \left( \frac{2 \pi}{\lambda_c} \epsilon  \right) \right].
\label{Eq10}
\end{split}
\end{equation}
The sensor signal is still proportional to a sine function of the wavefront step height, but modulated by a cosine distribution (because of the simultaneous two lines), itself modulated by a Gaussian distribution (because of the broadband nature of the two lines). The periodic nature of the sine is no longer a problem because it decays as the wavefront step height error increases (as for the coherence method). This corresponds to the coherence signature that can be exploited. Similarly to the coherence length defined in Eq. \ref{Eq3} for the coherence method, a coherence length, denoted $\mathcal{L}_c$, can be associated with the Gaussian function introduced in Eq. \ref{Eq10}, and is given by
\begin{equation}
\mathcal{L}_c = \frac{\lambda_c^2}{\delta \lambda}.  
\label{Eq11}
\end{equation}
Finally, Eq. \ref{Eq9} can be expressed as function of the coherence length as
\begin{equation}
\sigma_c =  \frac{2 \pi}{\sqrt{8 \ln 2}}\times  \frac{1}{\mathcal{L}_c}.
\label{Eq9b}
\end{equation}

\subsection{Practical considerations and required technology}
\begin{figure}[!htbp]
\includegraphics[scale=0.23]{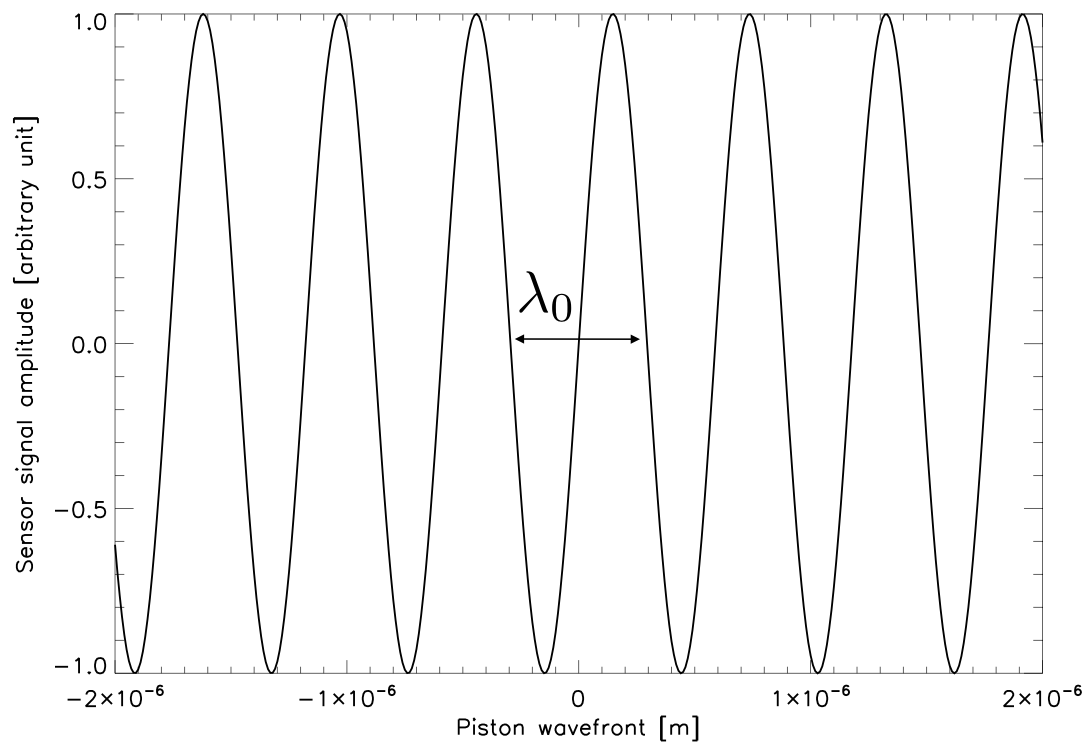}
\includegraphics[scale=0.23]{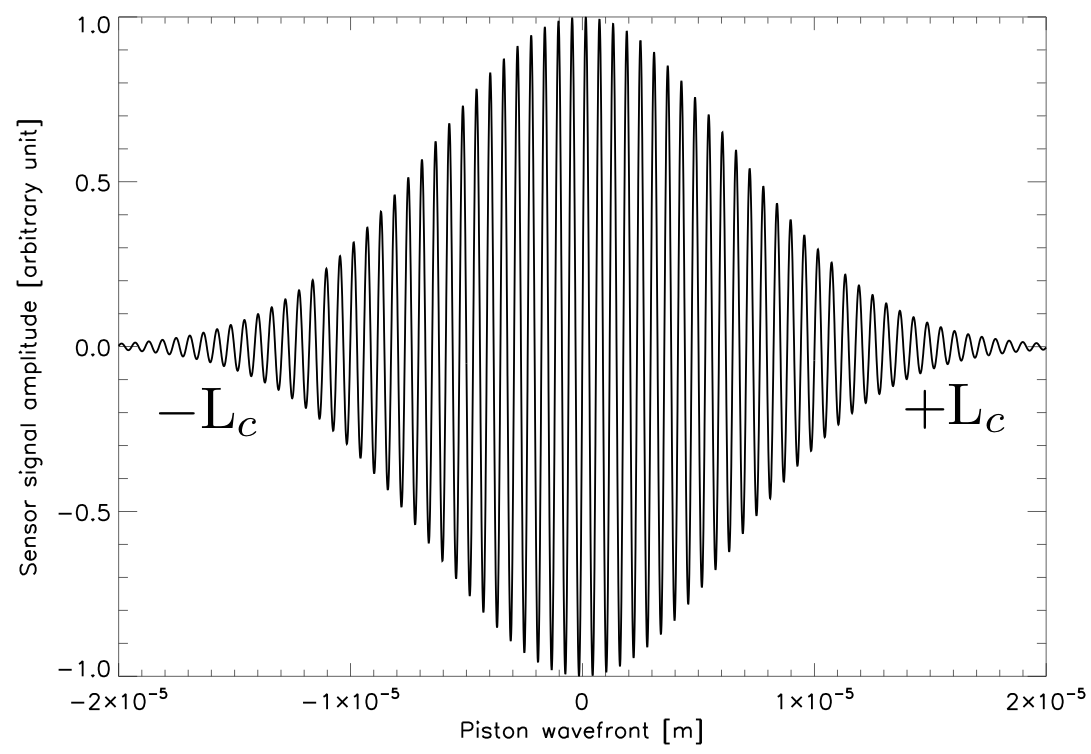}
\includegraphics[scale=0.23]{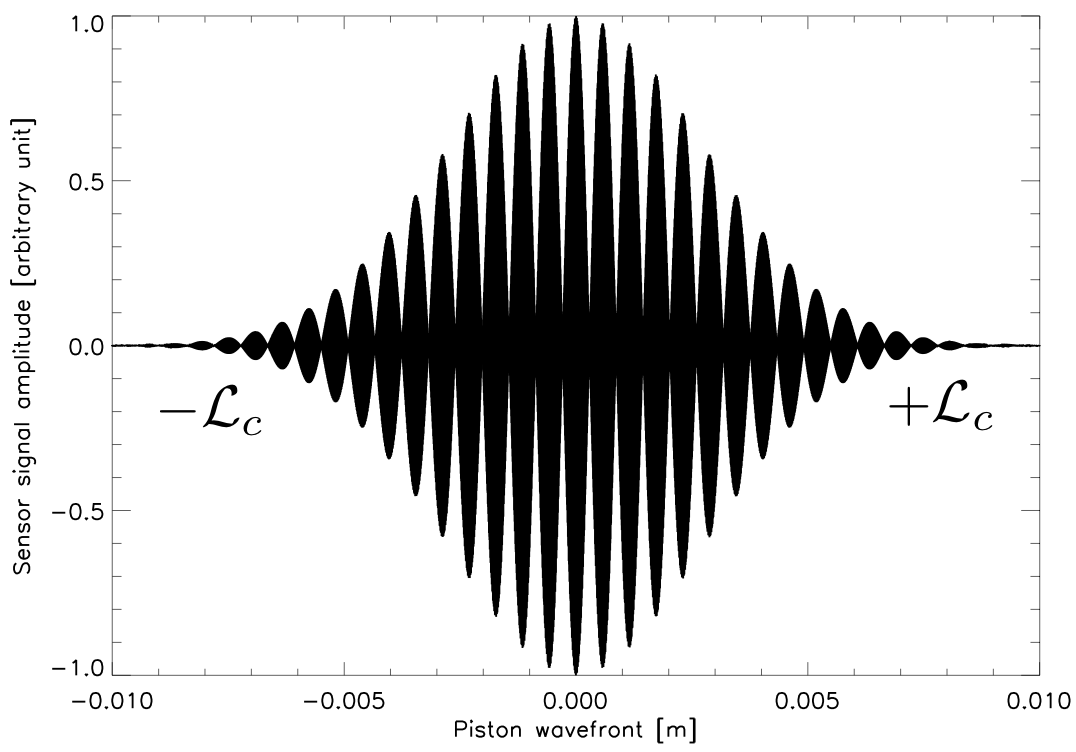}
\caption{Cophasing sensor signal as a function of the wavefront step: in the monochromatic regime (top) intrinsically limited by the $\pi$-ambiguity, using the coherence method (middle) with an extended capture range of $\pm \mathrm{L_c}$, and using the doublet-wavelength coherence method (bottom) with a broad capture range limited to $\pm \mathcal{L}_c$.}
\label{FIGURE}
\end{figure}

Figure \ref{FIGURE} compares a cophasing sensor signal in three different states: (a) within the monochromatic domain at 589 $nm$ (top panel), (b) using the coherence method at 589 $nm$ with $\delta \lambda_F = 20$ $nm$ (middle panel), and finally (c) using  the DWCT assuming $\lambda_1=589.0$ $nm$ and $\lambda_2=589.6$ $nm$, with $\Delta \lambda=0.6$ $nm$ and $\delta \lambda=0.05$ $nm$ (bottom panel). The increase in capture range delivered by the coherence method over the monochromatic regime, from (a) to (b), and by the DWCT over the coherence method, from (b) to (c), are readily observable. The DWCT gives access to a range of wavefront step heights on the order of a few $mm$ that is
not accessible with current techniques. We note that the general description of the DWCT has been verified with the self-coherent camera-phasing sensor \citep[SCC-PS, ][]{Potiron16} by means of numerical simulations.

Since $\mathrm{L_c}$ is inversely proportional to the filter bandpass $\delta \lambda_F$ and $\mathcal{L}_c$ is inversely proportional to the line width $\delta \lambda$, it is trivial to see that $\mathcal{L}_c \gg \mathrm{L_c}$ because $\delta \lambda_F \gg \delta \lambda$. This leads to a gain of at least a factor of 100 in the capture range of the DWCT over the coherence method. Now, the precision at the best-fit alignment must be discussed. 
The coherence method indeed changes the spectral bandwidth of the optical filter (by changing the filter) during operation to improve the accuracy of the best-fit alignment. The measurements start with a narrow bandwidth filter ($\sim$10 nm) covering a broad range of pistons at the expense of limited accuracy (the highest coherence corresponding to the best-fit alignment, i.e., the central oscillation of the sine term is diluted in the numerous oscillations contained inside the exponential envelope, see Fig. \ref{FIGURE} middle panel). The bandwidth of the optical filter is then increased ($\sim$100 nm) to improve the measurement precision
of the best-fit alignment. 
Exploiting the narrowband nature of a single line LGS, or in
other words, applying the coherence method with a single- and mono-line LGS, would then improve the capture range, but at the expense of the precision of the coherence method. A single-line LGS opens access to step (1), but cannot replace the coherence method as a step (2).

With the DWCT, the doublet-line introduces an additional cosine term to the sensor signal (see Eq. \ref{Eq10}). 
Figure \ref{FIGURE} (bottom panel) illustrates the three components contained in the sensor signal: the exponential envelope, the cosine term (black lobes), and the sine term included in each lobe of the cosine term (here undistinguishable). 
This specific cosine signature in the signal is fundamental to identifying the central one and thereby guarantees the precision of the best-fit alignment. Around its position, individual segments can then be successively poked by varying wavefront step heights to scan over the sine term where the final best-fit position can be found with accuracy. 
The width of these lobes are on the order of a tenth of $mm$, therefore we can expect to reach $\mu m$ or ultimately sub-$\mu m$ accuracy. While the first level would reduce the piston error to the capture range of step (2), the second level would eventually lead to step (3).

A practical implementation of the DWCT could take form either as the simultaneous use of two LGSs lasing at two different sodium D lines, or using a single LGS lasing on the D1 and D2 lines simultaneously. While the first alternative is certainly more convenient, successful laboratory demonstration of lasing on the D lines of sodium has been successfully demonstrated \citep{Hewitt2012}. 

\section{Conclusion}
\label{conclusion}
The simplified description given in this paper is not thoroughly satisfactory because LGS systems are rather complex. Nonetheless,
the preliminary analysis of the DWCT with LGSs suggests that the approach is worthy of interest. 
By using both the coherence and the double-line properties, the DWCT simultaneously grants access to steps (1) and (2) in one single interferogram and thus represents an enhancement of the coherence method.
Advantages are not only the broad capture range, but also the inherent stability of the method, because the response coherence curve is Gaussian, and not periodic. 
The method is well suited to the re-integration of segments (or integration of spare segments) on a daily basis that return from re-coating. 
The DWCT improves the phasing optics capture paradigm to the
millimetric domain and makes the man-made mechanical pre-phasing step redundant 
This will accelerate the coarse-phasing step before the final fine-phasing step. If that lasing on multiple sodium D lines can be implemented, then the DWCT is a powerful tool that can also be used with phasing sensor designs.
Applications to multi-aperture interferometric arrays still need to be explored. 

\begin{acknowledgements}
P. Janin-Potiron is supported by Airbus Defense and Space (Toulouse, France) and the R\'egion PACA (Provence Alpes C\^ote d\'{}Azur, France). We thank the anonymous referee for the useful comments on the manuscript.  
\end{acknowledgements}

\bibliography{biblio}

\end{document}